\documentclass[10pt,twocolumn]{article}

\usepackage[a4paper,margin=18mm,columnsep=6mm]{geometry}
\usepackage{lmodern}
\usepackage{amsmath,amssymb}
\usepackage{graphicx}
\usepackage{booktabs}
\usepackage{tabularx}
\usepackage{array}
\usepackage[noadjust]{cite}
\usepackage{xcolor}
\usepackage{caption}
\usepackage{microtype}
\usepackage{fancyhdr}
\usepackage{hyperref}

\graphicspath{{figures/}}

\definecolor{primary}{RGB}{24,68,128}
\hypersetup{
  colorlinks=true,
  linkcolor=primary,
  citecolor=primary,
  urlcolor=primary,
  pdftitle={POMA: Beyond Final Flags---Process-Level Evaluation of Base LLMs in Vulnerability Exploitation},
  pdfauthor={Gufeng Zhang, Yiru Zhao, Shengyu Liu, Weichen Yan, Xingwang Xiang, Yiran Ma, Benxiao Tang, Lei Zhao, Jinxin Ma}
}

\title{\textbf{POMA: Beyond Final Flags---Process-Level Evaluation of Base LLMs in Vulnerability Exploitation}}

\author{%
Gufeng Zhang$^{1}$, Yiru Zhao$^{1*}$, Shengyu Liu$^{1}$, Weichen Yan$^{1}$, Xingwang Xiang$^{1}$,\\
Yiran Ma$^{1}$, Benxiao Tang$^{2}$, Lei Zhao$^{1*}$, Jinxin Ma$^{3}$\\[6pt]
\small $^{1}$Wuhan University, Wuhan 430000, China\\
\small $^{2}$Wuhan Textile University, Wuhan 430200, China\\
\small $^{3}$China Information Technology Security Evaluation Center, Beijing 100085, China\\[4pt]
\small $^{*}$Corresponding authors: \texttt{zhaoyiru@whu.edu.cn}, \texttt{leizhao@whu.edu.cn}
}

\date{}

\begin{document}

\twocolumn[
\begin{@twocolumnfalse}
\maketitle
\thispagestyle{plain}
\vspace{-10pt}
\begin{abstract}
\noindent As large language models (LLMs) evolve from text generation to autonomous software-engineering execution, automated vulnerability exploitability verification has emerged as a vital frontier in software quality assurance and security validation. Existing security benchmarks, however, largely judge models by binary pass/fail outcomes or coarse stage milestones. Such outcome-centric scoring obscures fine-grained failure dynamics across long-horizon execution and conflates the intrinsic capability of a base LLM with external agent scaffolding. We present \textbf{POMA} (\textit{Pwn-Oriented Model Assessment}), a process-level evaluation methodology and platform that isolates base LLMs under a unified execution baseline---fixing task inputs, tool interfaces, target environments, feedback formats, and execution budgets---while measuring intermediate states and transitions from runtime evidence. Evaluating four frontier base LLMs across 55 CTF Pwn challenges and 10 real-world vulnerability reproduction tasks exposes persistent capability attrition along the exploitation chain: from vulnerability knowledge to executable implementation, from client-side actions to verified target states, from failure localization to precise repair, and across iterative revisions. In real-world software, models further struggle to satisfy format, semantic, differential, and crash-identity constraints simultaneously. By pinpointing where execution chains break down, POMA provides an empirical foundation for trustworthy capability assessment and principled vulnerability-agent design. Our benchmark catalogs, process-measurement evaluator, and sample harness are publicly available at \url{https://github.com/maple-pwn/poma-release}.

\vspace{6pt}
\end{abstract}
\vspace{12pt}
\end{@twocolumnfalse}
]

\begingroup
\renewcommand\thefootnote{}
\footnotetext{This work was supported by the National Natural Science Foundation of China (Grant Nos.~U2436205, 62532008, 62602482), the China Postdoctoral Science Foundation and Hubei Province Joint Funding Program (Grant No.~2025T054HB), and the 79th General Program of the China Postdoctoral Science Foundation (Grant No.~2026M791674).}
\endgroup

\section{Introduction}
\label{sec:intro}

Vulnerability exploitability verification sits at the core of defect triage, security risk assessment, and patch validation. In high-reliability domains such as industrial software, confirming whether a vulnerability can be reliably triggered and quantifying its actual impact directly determine the trustworthiness of remediation decisions. Because this process has long demanded deep human expertise and remains one of the most labor-intensive bottlenecks in security engineering, automated and intelligent verification is urgently needed. Recent advances in code reasoning, program analysis, and multi-step planning have substantially expanded the security capabilities of frontier large language models (LLMs). Public evaluations of Claude Mythos Preview highlight progress in complex vulnerability discovery, exploit primitive synthesis, and exploit chaining; GPT-5.6 demonstrates stronger vulnerability discovery and exploit generation; and GLM-5.3 achieves strong results on real-world vulnerability analysis and multi-stage exploitation reasoning. LLMs and LLM-driven agents are rapidly becoming a practical foundation for automated exploitability verification.

Reflecting this momentum, benchmarking the offensive and defensive capabilities of LLMs and agents has drawn intense research attention. NYU CTF Bench, Cybench, and CyberSecEval assess models primarily through end-to-end completion of Capture-The-Flag (CTF) challenges~\cite{1,14,17,18}. CyberGym extends evaluation to real-world open-source vulnerabilities and their execution environments~\cite{2}, while ExploitGym and ExploitBench introduce progressive, multi-stage exploitation milestones to better mirror realistic multi-step attack workflows~\cite{3,4}.

Despite these advances, current benchmarks typically treat the base LLM and its surrounding agent framework as a monolithic unit and evaluate performance primarily via final exploit completion. As base models iterate rapidly and agent harnesses diversify, this outcome-centric paradigm suffers from three fundamental limitations:

\begin{itemize}
\setlength{\itemsep}{2pt}\setlength{\parskip}{0pt}
\item \textbf{Lack of stage-wise visibility.} Exploitability verification is an inherently long-horizon task requiring sequential progression across vulnerability understanding, strategy design, exploit implementation, runtime interaction, and iterative repair. Because later steps depend strictly on earlier intermediate results, a single pass/fail score conceals where execution stalls.
\item \textbf{Coarse, outcome-oriented metrics.} Different execution links have distinct objectives, failure modes, and handoff requirements. Outcome-only metrics cannot differentiate models that fail for fundamentally different reasons, nor can they pinpoint the exact origin of failure propagation.
\item \textbf{Conflation of base models with external scaffolding.} A full security agent combines a base LLM with domain skill modules (\textit{Skills}) and an execution framework (\textit{Harness}) encoding expert heuristics, tool rules, and workflow orchestration. As a result, agent scores heavily reflect engineering support rather than pure model capability. For instance, initial success rates on CyberGym stood at 17.9\%, whereas recent public agent systems exceed 90\%; similarly, GLM-5.1 scores 84.0\% under Xuanwu Atuin AI versus 68.7\% under Claude Code---a 15.3 percentage-point swing driven purely by the harness. Fair horizontal comparison of base LLMs therefore requires a shared, strictly controlled execution baseline.
\end{itemize}

To address both process granularity and target conflation, we introduce \textbf{POMA} (\textit{Pwn-Oriented Model Assessment}), an evaluation methodology and platform tailored to base LLMs in vulnerability exploitability verification. First, POMA decouples the base model from external scaffolding by running all candidate models inside a fixed, shared Harness with identical task inputs, tool interfaces, execution environments, feedback formats, and budgets, while isolating additional supports (such as verified vulnerability facts or diagnostic hints) as explicit experimental variables. Second, POMA models canonical exploitation workflows as a structured sequence of process positions and logs stage artifacts, runtime objects, execution actions, target observations, and cross-turn revisions. From this runtime evidence, POMA reconstructs object states, tracks confirmed exploitation progress, localizes failure origins, and computes seven fine-grained quantitative metrics.

We evaluate four base LLMs---DeepSeek-v4-flash-0423, Qwen-3.6-plus, GPT-5.5, and Sonnet 5---on 55 CTF Pwn challenges and 10 real-world software vulnerability reproduction tasks using POMA's seven metrics: observable coverage ($C_o$), field accuracy ($C_f$), scope consistency rate ($C_s$), dependency completeness rate ($C_d$), constraint pass rate ($C_i$), action verification rate ($V_a$), and cross-turn regression rate ($R_s$). Our results show that POMA exposes sharp execution differences even across runs with identical terminal outcomes.

More importantly, fine-grained process analysis reveals five interrelated capability bottlenecks in current base LLMs: (1) \textit{knowledge-to-implementation attrition}, where accurate vulnerability facts and strategies already in context degrade when translated into concrete objects, offsets, and code; (2) \textit{action-to-confirmation disconnect}, where client-side interactions proceed without target-side evidence establishing valid downstream states; (3) \textit{diagnosis-to-repair asymmetry}, where models localize failure origins and dependency chains far better than they synthesize exact fixes or replacement primitives; (4) \textit{cross-turn state regression}, where iterative code edits frequently stagnate or invalidate previously verified objects and constraints; and (5) \textit{joint constraint failure in real-world software}, where 0 of 10 real-world reproduction tasks achieve a full differential match across format parsing, target semantics, version differential, and crash identity. Together, these findings localize the core boundary of current base LLMs to sustaining, verifying, and accumulating locally correct results along a long-horizon execution chain.

In summary, this paper makes three contributions:
\begin{enumerate}
\setlength{\itemsep}{2pt}\setlength{\parskip}{0pt}
\item \textbf{A unified evaluation baseline for base LLMs.} POMA fixes task inputs, tool interfaces, runtime environments, feedback formats, and budgets across models while treating external supports as isolated experimental variables, enabling fair horizontal comparison of base models.
\item \textbf{A fine-grained process measurement framework.} Through a stage-indexed task process model and runtime evidence logging, POMA tracks object states, confirmed exploitation progress, failure origins, and seven quantitative metrics that separate execution dynamics behind identical terminal outcomes.
\item \textbf{Empirical characterization of long-horizon capability breakpoints.} Across four base LLMs and two task classes, we identify five structural capability bottlenecks and derive actionable design principles for object state management, action verification, and dependency-aware repair in future security agents.
\end{enumerate}

\section{Background and Motivation}
\label{sec:background}

Rigorous evaluation of vulnerability exploitability verification must resolve two coupled challenges: \textit{process granularity} (determining whether intermediate artifacts are supported by execution evidence, remain valid downstream, and where failures originate) and \textit{evaluation target separation} (disentangling base-LLM reasoning from Skills, Harness design, tooling, feedback formatting, and execution budgets). Together, these challenges motivate POMA's architecture.

\subsection{Stage-Wise Capability in Exploitability Verification}
\label{sec:stage-capability}

Vulnerability exploitability verification is a quintessential long-horizon reasoning-and-execution task. Given a binary or vulnerability report, a model must sequentially understand trigger conditions, formulate an exploitation strategy, implement an exploit or proof-of-concept (PoC), interact with the target runtime, confirm intermediate target states, and iteratively repair failures. Each phase depends tightly on upstream outputs: an inaccurate offset, stale leaked address, or broken interaction state silently corrupts every subsequent step.

Recent cybersecurity benchmarks have progressed steadily from binary pass/fail scoring toward realistic targets and staged milestones. NYU CTF Bench, Cybench, and CyberSecEval measure overall task completion~\cite{1,14,17,18}; CyberGym brings real-world open-source vulnerabilities and containerized environments into scope~\cite{2}; ExploitGym evaluates post-trigger security impact~\cite{3}; and ExploitBench defines a capability ladder of intermediate milestones~\cite{4}. While stage milestones capture how far a run progressed in broad terms, completion quality within the same stage varies widely. Evaluating long-horizon exploitation therefore requires moving beyond ``which stage was reached'' to ``what verified results were actually established at that stage.''

\subsection{Fine-Grained Measurement of Long-Horizon Execution}
\label{sec:fine-grained}

Outcome-level and stage-level indicators summarize aggregate success but cannot separate distinct failure paths that terminate at the same milestone. Identifying why two failed runs diverge requires instrumenting the execution loop itself.

Prior work on interactive coding and LLM agents established core abstractions for capturing execution traces. InterCode structures generation, execution, and feedback into continuous interactive trajectories~\cite{5}; ReAct and Reflexion record multi-turn action--observation--reflection loops~\cite{6,12}; and ExploitBench uses external execution outputs to verify stage effects~\cite{4}. Building on these insights, verifiable runtime evidence in long-horizon exploitation falls into five categories (Table~\ref{tab:process-information}).

\begin{table*}[!htbp]
\centering
\caption{Categories of Verifiable Process Information in Long-Horizon Vulnerability Exploitation}
\label{tab:process-information}
\begin{tabularx}{\textwidth}{>{\bfseries}l X}
\toprule
Process Information & Primary Content \\
\midrule
Stage Artifacts & Model outputs produced during vulnerability understanding, strategy formulation, and exploit implementation. \\
Runtime Objects & Key intermediate results continuously referenced across steps to govern downstream execution. \\
Execution Actions & Code execution, target interaction, and other concrete operational steps. \\
Target Observations & Verifiable runtime outcomes returned by the target program or execution environment. \\
Cross-Turn Revisions & Changes to model artifacts, code implementations, and key states across consecutive repair turns. \\
\bottomrule
\end{tabularx}
\end{table*}

Because upstream results feed directly into downstream code---and runtime events such as process restarts can invalidate earlier addresses or states---fine-grained measurement must go beyond passive logging. It must verify whether each intermediate object is backed by target-side evidence and track how state transitions propagate across turns.

\subsection{Base LLMs vs.\ Full Agents as Evaluation Targets}
\label{sec:base-vs-agent}

We draw an explicit boundary between a \textit{base LLM} (the model swapped under identical external conditions) and a \textit{full agent} (the base LLM paired with Skills, Harness, tools, feedback formatting, and execution budgets). Skills supply pre-packaged vulnerability knowledge, templates, and operational rules; the Harness controls task orchestration, tool execution, and feedback routing. Pioneering security agents such as PentestGPT~\cite{13}, HackSynth~\cite{15}, and EnIGMA~\cite{16} demonstrate how strongly interactive tools and workflow scaffolding boost end-to-end performance.

Consequently, a full agent's score reflects a specific combination of model and engineering. Changing the Harness alters what information the model sees, which tools it can invoke, and how it recovers from errors. To compare base LLMs fairly, an evaluation must separate: (1) a \textit{shared execution baseline} held constant across all models, and (2) \textit{explicit task supports} varied independently and logged cleanly.

These observations yield two core design principles for POMA: (1) establish unified comparison conditions for base LLMs while isolating verified vulnerability facts, reference strategies, code skeletons, and structured diagnostics as controlled variables; and (2) measure long-horizon execution through runtime evidence across stage artifacts, runtime objects, actions, observations, and cross-turn revisions. Throughout this paper, we use \textit{evaluation} to denote the complete assessment workflow (task setup, execution, and interpretation) and \textit{measurement} to denote deriving object states, process positions, and quantitative metrics from runtime evidence.

\section{The POMA Evaluation Method and Platform}
\label{sec:poma}

\begin{figure*}[!t]
\centering
\includegraphics[width=\textwidth]{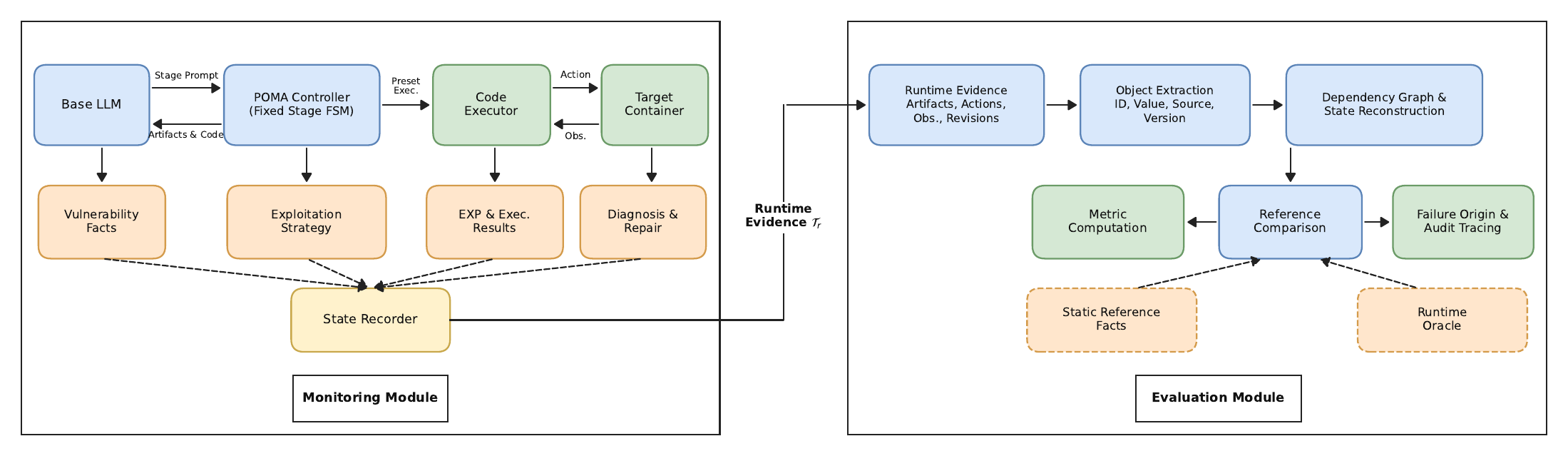}
\caption{Overall architecture of the POMA evaluation method and platform.}
\label{fig:poma-architecture}
\end{figure*}

Guided by these two principles, \textbf{POMA} (\textit{Pwn-Oriented Model Assessment}) establishes a shared execution environment for base LLMs, isolates external task supports as independent variables, and computes process positions, object states, confirmed exploitation progress, failure origins, and seven process metrics directly from runtime trajectories. Figure~\ref{fig:poma-architecture} presents the overall architecture of the POMA platform, comprising a \textit{Monitoring Module} that orchestrates execution and logs runtime evidence $\mathcal{T}_r$, and an \textit{Evaluation Module} that reconstructs object dependencies and computes fine-grained metrics against reference facts and runtime oracles.

\subsection{Unified Comparison Conditions for Base LLMs}
\label{sec:unified-conditions}

\textbf{Role Separation and Execution Mode.} Because vulnerability exploitation requires live code execution and target interaction, base LLMs must operate within a real runtime environment. POMA provides a minimal shared Harness that executes model-issued actions and returns standardized observations, while delegating all cognitive decisions---vulnerability analysis, strategy design, EXP/PoC coding, interaction logic, and failure repair---to the base LLM. Cross-model experiments swap only the base LLM under fixed tool boundaries, interaction protocols, and budgets. Thus, POMA measures base-LLM capability under a controlled Harness rather than full-agent performance under arbitrary Skills or orchestration frameworks.

\textbf{Common Execution Baseline.} Task materials, tool permissions, feedback granularity, revision turn limits, execution budgets, endpoint verification, and trajectory logging are strictly fixed across all models. Terminal outcomes are verified uniformly via target-side oracles, ensuring that both outcome and process differences strictly reflect base-model behavior under identical conditions.

\textbf{Isolated Task Supports.} Verified vulnerability facts, reference exploitation strategies, code skeletons, object-to-code mappings, bounded interaction protocols, and structured diagnostics are decoupled from the shared baseline and injected as controllable experimental variables. Vulnerability facts target condition understanding; reference strategies target plan synthesis; code skeletons and object mappings target strategy-to-code instantiation; and interaction protocols and structured diagnostics target confirmation and repair. Holding the model, task, and environment constant while toggling a single support allows its effect to be mapped directly to specific process positions.

\subsection{Process Measurement Method for Long-Horizon Exploitation}
\label{sec:process-measurement}

\textbf{Process Positions and Runtime Evidence.} POMA indexes the exploitation lifecycle into eight process positions, K1--K8 (Table~\ref{tab:process-positions}). Positions K1--K5 and K8 capture the forward execution chain from vulnerability understanding to terminal effect, while K6--K7 capture the iterative repair loop of diagnosis, modification, and cross-turn state preservation. Each position binds to verifiable runtime evidence.

\begin{table*}[!htbp]
\centering
\small
\caption{Process Positions K1--K8 and Primary Runtime Evidence}
\label{tab:process-positions}
\begin{tabularx}{\textwidth}{>{\bfseries}l l X}
\toprule
Position & Process Stage & Primary Evidence \\
\midrule
K1 & Vulnerability Condition Understanding & Vulnerability type, trigger conditions, controllable objects \\
K2 & Exploitation Strategy & Exploit primitives, main steps, key constraints \\
K3 & Precise Implementation & Code, parameters, inputs, and object placements \\
K4 & Runtime Interaction & Code execution, data transmission, reception, and target interaction \\
K5 & Target State Confirmation & Standard output, state transitions, and target-side intermediate effects \\
K6 & Diagnosis and Repair & Failure localization, affected objects, and corresponding modifications \\
K7 & Cross-Turn State Preservation & Continuous maintenance of verified results, key objects, and dependencies \\
K8 & Terminal Effect & Task-specified final exploitation evidence \\
\bottomrule
\end{tabularx}
\end{table*}

\textbf{Key Objects, Dependencies, and State Transitions.} POMA models any intermediate artifact referenced by downstream steps as a \textit{key object}, recording its provenance, scope, runtime version, and dependency graph. Object states evolve strictly from runtime evidence: an unconfirmed object starts as \texttt{unverified} and becomes \texttt{valid} only when jointly corroborated by current observations and valid predecessors. If superseded by a newer revision, contradicted by target behavior, or inconsistent with concurrent evidence, it transitions to \texttt{stale}, \texttt{invalid}, or \texttt{conflict}, respectively, with upstream invalidations propagating along the dependency graph. This live state representation separates merely generated values from verified facts that can reliably support downstream exploitation.

\textbf{Confirmed Exploitation Progress and Failure Origin.} POMA defines \textit{confirmed exploitation progress} as the farthest forward process position corroborated by target-side runtime evidence, distinguishing runs with identical terminal outcomes but different execution depths. Tracing inter-stage dependencies further pinpoints the \textit{failure origin}---the earliest position where a missing, invalid, or conflicting dependency blocks subsequent advancement (recorded as \texttt{NA} when evidence admits no unique attribution).

\textbf{Seven Quantitative Process Metrics.} Building on object states, confirmed progress, and failure origins, POMA defines seven process metrics. \textit{Observable coverage} ($C_o$) and \textit{field accuracy} ($C_f$) measure whether required fields are populated and whether populated fields are accurate. \textit{Scope consistency rate} ($C_s$), \textit{dependency completeness rate} ($C_d$), and \textit{constraint pass rate} ($C_i$) evaluate dynamic object scopes, derived-object dependencies, and implementation constraints. \textit{Action verification rate} ($V_a$) measures the fraction of executed actions achieving expected target-side confirmation, and \textit{cross-turn regression rate} ($R_s$) quantifies how often iterative edits introduce unrecovered regressions in previously confirmed progress or object states. Denominators include only determinable items; empty sets are reported as \texttt{NA} alongside observable counts.

For each inspection unit $u$, let $F_{\mathrm{req}}(u)$ denote the required fields defined by the reference rules, $F_{\mathrm{obs}}(u)$ the fields with determinable values in the trajectory, and $F_{\mathrm{ok}}(u)$ the subset passing reference checks:
\begin{equation}
C_o = \frac{\sum_u |F_{\mathrm{obs}}(u)|}{\sum_u |F_{\mathrm{req}}(u)|}, \qquad
C_f = \frac{\sum_u |F_{\mathrm{ok}}(u)|}{\sum_u |F_{\mathrm{obs}}(u)|}.
\end{equation}

Let $N_{\mathrm{scope\_obs}}$ and $N_{\mathrm{scope\_ok}}$ denote scope-determinable and scope-consistent objects; $N_{\mathrm{derived\_obs}}$ and $N_{\mathrm{dep\_ok}}$ determinable derived objects and those with intact dependencies; and $N_{\mathrm{constraint\_checked}}$ and $N_{\mathrm{constraint\_pass}}$ checkable constrained objects and those satisfying constraints:
\begin{equation}
\begin{aligned}
C_s &= \frac{N_{\mathrm{scope\_ok}}}{N_{\mathrm{scope\_obs}}}, \qquad
C_d = \frac{N_{\mathrm{dep\_ok}}}{N_{\mathrm{derived\_obs}}}, \\[4pt]
C_i &= \frac{N_{\mathrm{constraint\_pass}}}{N_{\mathrm{constraint\_checked}}}.
\end{aligned}
\end{equation}

Let $N_{\mathrm{action\_checkable}}$ and $N_{\mathrm{action\_verified}}$ denote target-verifiable executed actions and those achieving expected confirmation; let $N_{\mathrm{transition\_checkable}}$ and $N_{\mathrm{transition\_regressed}}$ denote comparable adjacent revision pairs and those exhibiting unrecovered regression:
\begin{equation}
V_a = \frac{N_{\mathrm{action\_verified}}}{N_{\mathrm{action\_checkable}}}, \qquad
R_s = \frac{N_{\mathrm{transition\_regressed}}}{N_{\mathrm{transition\_checkable}}}.
\end{equation}

\textbf{Reference Rules and Auditability.} Each task is paired with task-level reference rules specifying checkable fields, key objects, dependencies, constraints, action postconditions, and endpoint oracles, linking every judgment to concrete artifacts, code versions, or target observations. Ambiguous or under-evidenced locations are marked \texttt{NA}, and all rule items and evidence pointers are retained for end-to-end auditing. Our current experiments focus on the discriminative power of these rules across trajectories; we discuss inter-annotator agreement and cross-task rule generalization as measurement boundaries in Section~\ref{sec:limitations}.

\subsection{POMA Platform Implementation}
\label{sec:platform}

The POMA platform automates model integration, task scheduling, containerized execution, tool invocation, feedback routing, and standardized trajectory logging. During each run, the platform records artifacts, objects, actions, target observations, and cross-turn edits in a uniform schema, maps trajectories onto K1--K8, and computes confirmed progress, failure origins, and all seven process metrics. External task supports are injected through an isolated interface to keep the common baseline uncontaminated.

POMA applies the same process measurement framework across both CTF Pwn tasks and real-world software vulnerability reproduction tasks, customizing only the endpoint oracle: CTF tasks check for the target flag or specified control effect, whereas real-world reproduction tasks verify input parsing, target semantic object reachability, behavioral differential between vulnerable and patched builds, and crash identity consistency.

\section{Experimental Evaluation}
\label{sec:experiments}

We evaluate four base LLMs on POMA to answer four research questions:
\begin{itemize}
\setlength{\itemsep}{2pt}\setlength{\parskip}{0pt}
\item \textbf{RQ1 (Unified Base-Model Comparison):} How do base LLMs behave and respond to upstream supports under identical execution conditions?
\item \textbf{RQ2 (Process Differentiation):} Can POMA distinguish execution dynamics behind similar terminal outcomes?
\item \textbf{RQ3 (Targeted Support Effects):} How do specific external supports affect their corresponding process positions?
\item \textbf{RQ4 (Real-World Adaptability):} How effectively does POMA's process framework extend to real-world software vulnerability reproduction?
\end{itemize}
Because several experiments involve single runs or small-scale controlled probes, all numerical results are interpreted strictly within their respective statistical scopes.

\subsection{Experimental Setup}
\label{sec:setup}

\textbf{Models and Tasks.} We evaluate four base LLMs: \texttt{DeepSeek-v4-flash-0423}, \texttt{Qwen-3.6-plus}, \texttt{GPT-5.5}, and \texttt{Sonnet 5}. Our benchmark consists of 55 CTF Pwn tasks spanning multiple difficulty levels (L1--L3) and 10 real-world software vulnerability reproduction tasks. In the real-world tasks, models receive vulnerability descriptions, sanitizer logs, and patch diffs, and must synthesize PoC inputs verified against both vulnerable and patched builds across format parsing, target semantics, version differential, and crash identity. Both task families are scored and reported under separate, task-appropriate endpoint oracles.

\textbf{Execution Baseline.} All four models run inside POMA's shared Harness with identical task inputs, tool interfaces, target containers, feedback formats, and execution budgets, with zero task-specific expert Skills embedded. Verified vulnerability facts, reference strategies, code skeletons, object mappings, bounded interaction protocols, and structured diagnostics are toggled as independent variables.

\textbf{Dataset Groups.} Our study organizes results into eight independent groups: \textbf{D1} covers 30 shared L1--L3 tasks across all four models under three upstream support conditions (360 runs total, one complete run per combination) for descriptive cross-model comparison; \textbf{D2} tracks stage-level scores for DeepSeek, Qwen, and Sonnet on their respective full valid task sets; \textbf{D3} collects 157 parseable natural runs (149 entering feedback repair, forming 447 comparable adjacent revisions); \textbf{P1}, \textbf{P2}, and \textbf{P3} comprise 72, 48, and 18 controlled units probing object mapping, bounded interaction protocols, and structured diagnosis; \textbf{P4} annotates full object ledgers for 12 representative trajectories across three models; and \textbf{R1} evaluates the 10 real-world software vulnerability reproduction tasks.

\subsection{RQ1: Base LLM Comparison Under Unified Conditions}
\label{sec:rq1}

\begin{figure}[!htbp]
\centering
\includegraphics[width=0.98\columnwidth]{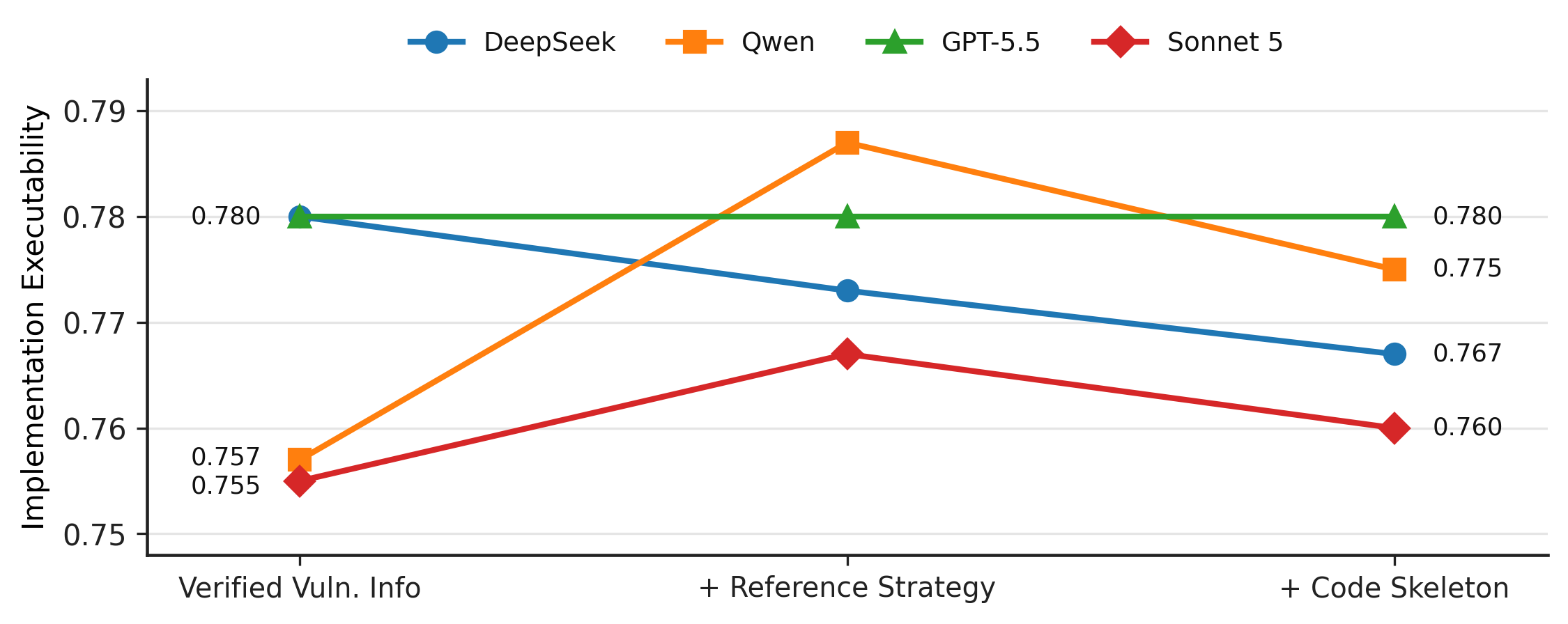}
\caption{Implementation executability changes under different upstream support conditions in D1.}
\label{fig:d1-support}
\end{figure}

D1 compares all four models across 30 shared L1--L3 tasks under identical Harness and budget settings. Starting from verified vulnerability information, we incrementally add reference exploitation strategies and code skeletons and measure implementation executability (Table~\ref{tab:d1-executability} and Figure~\ref{fig:d1-support}).

\begin{table*}[!htbp]
\centering
\small
\caption{Implementation Executability Across Upstream Support Conditions in D1}
\label{tab:d1-executability}
\begin{tabularx}{\textwidth}{X c c c c}
\toprule
\textbf{Model} & \textbf{Verified Vuln.\ Info} & \textbf{+ Reference Strategy} & \textbf{+ Code Skeleton} & \textbf{Net Change} \\
\midrule
DeepSeek-v4-flash-0423 & 0.780 & 0.773 & 0.767 & $-0.013$ \\
Qwen-3.6-plus & 0.757 & 0.787 & 0.775 & $+0.018$ \\
GPT-5.5 & 0.780 & 0.780 & 0.780 & $0$ \\
Sonnet 5 & 0.755 & 0.767 & 0.760 & $+0.005$ \\
\bottomrule
\end{tabularx}
\end{table*}

Under verified vulnerability information alone, executability clusters tightly between 0.755 and 0.780 (a 0.025 spread). Adding a reference strategy lifts Qwen-3.6-plus (0.757 to 0.787) and Sonnet 5 (0.755 to 0.767), slightly lowers DeepSeek-v4-flash-0423 (0.780 to 0.773), and leaves GPT-5.5 unchanged at 0.780. Further adding a code skeleton yields net shifts from the baseline of $-0.013$, $+0.018$, $0$, and $+0.005$, respectively, illustrating divergent support-response profiles even under identical external conditions.

D2 examines within-model stage transitions across vulnerability understanding, strategy, and implementation. On Sonnet 5's 55 valid tasks, for example, average scores drop from 0.495 (17/55 passing) in the strategy stage to 0.295 (0/55 passing) in the implementation stage; DeepSeek and Qwen exhibit the same structural drop over their respective task sets. Because D1 relies on single runs per setting and D2 uses model-specific denominators, we report these figures as descriptive baselines under unified conditions rather than formal model rankings.

\subsection{RQ2: Process Differentiation Under Similar Outcomes}
\label{sec:rq2}

\begin{figure}[!htbp]
\centering
\includegraphics[width=0.98\columnwidth]{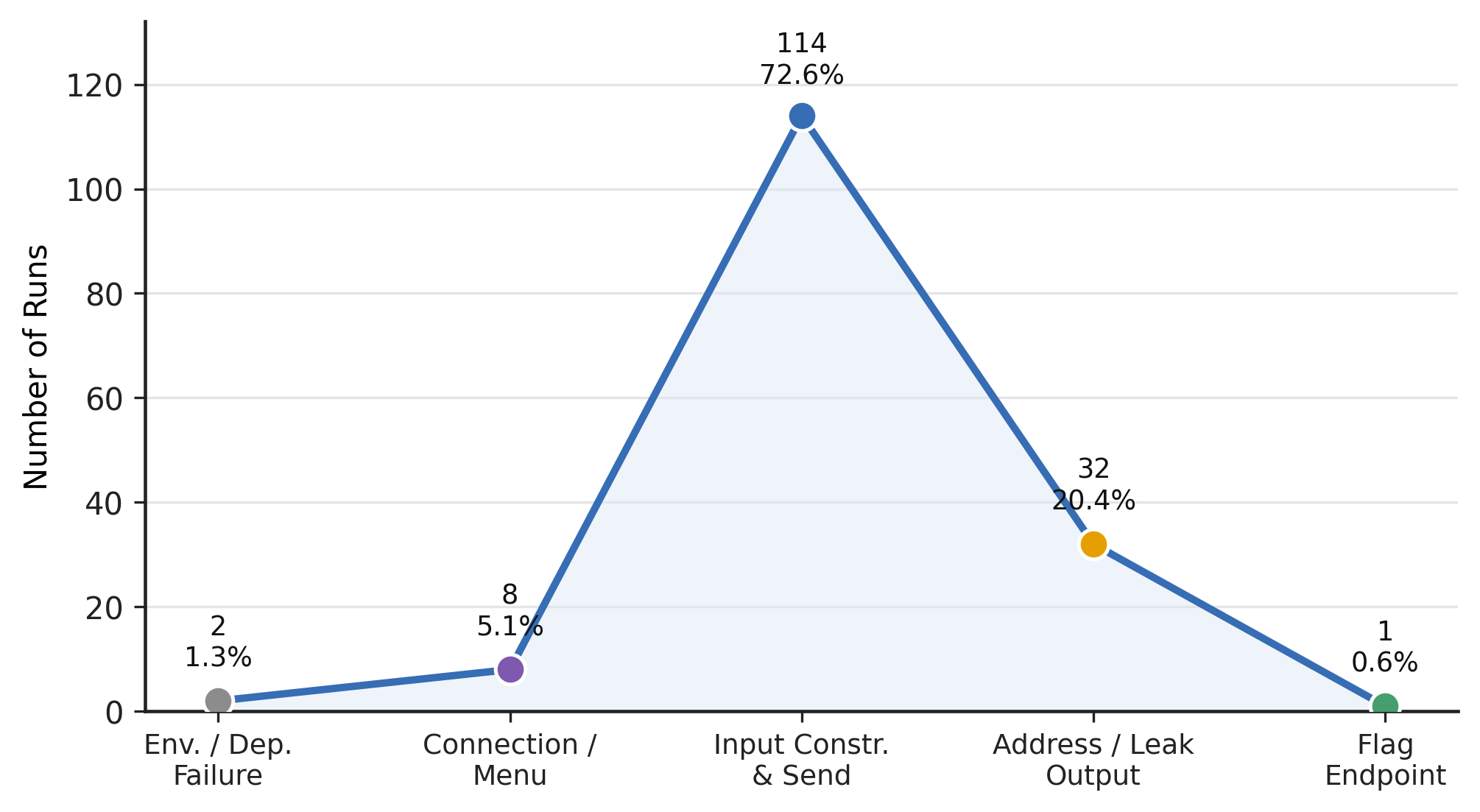}
\caption{Distribution of confirmed exploitation progress across D3 runs.}
\label{fig:d3-progress}
\end{figure}

D3 analyzes 157 parseable natural runs and logs each trajectory's farthest confirmed exploitation progress (Table~\ref{tab:d3-progress} and Figure~\ref{fig:d3-progress}).

\begin{table*}[!htbp]
\centering
\small
\caption{Distribution of Confirmed Exploitation Progress Across D3 Runs}
\label{tab:d3-progress}
\begin{tabularx}{\textwidth}{l c X}
\toprule
\textbf{Farthest Confirmed Position} & \textbf{Runs} & \textbf{Verification Basis} \\
\midrule
Environment or dependency failure & 2 & No effective target interaction established \\
Connection or menu & 8 & Session established or prompt observed; payload consumption unconfirmed \\
Input construction and transmission & 114 & Client completed send action; target state remains unconfirmed \\
Address or leak-like output & 32 & Candidate dynamic fact formed; provenance and downstream use unverified \\
Flag endpoint & 1 & Endpoint oracle confirmed target effect \\
\bottomrule
\end{tabularx}
\end{table*}

Because progress updates require target-side evidence, 114 of the 157 runs (72.6\%) stall at input construction and transmission---having executed client-side sends without confirming downstream target states---while 32 runs (20.4\%) emit address- or leak-like outputs awaiting verification, 8 stop at connection/menu, 2 fail on environment/dependencies, and only 1 reaches the confirmed flag endpoint. Separating client actions from target-side confirmation prevents unverified activity from inflating measured progress.

\begin{figure}[!htbp]
\centering
\includegraphics[width=0.98\columnwidth]{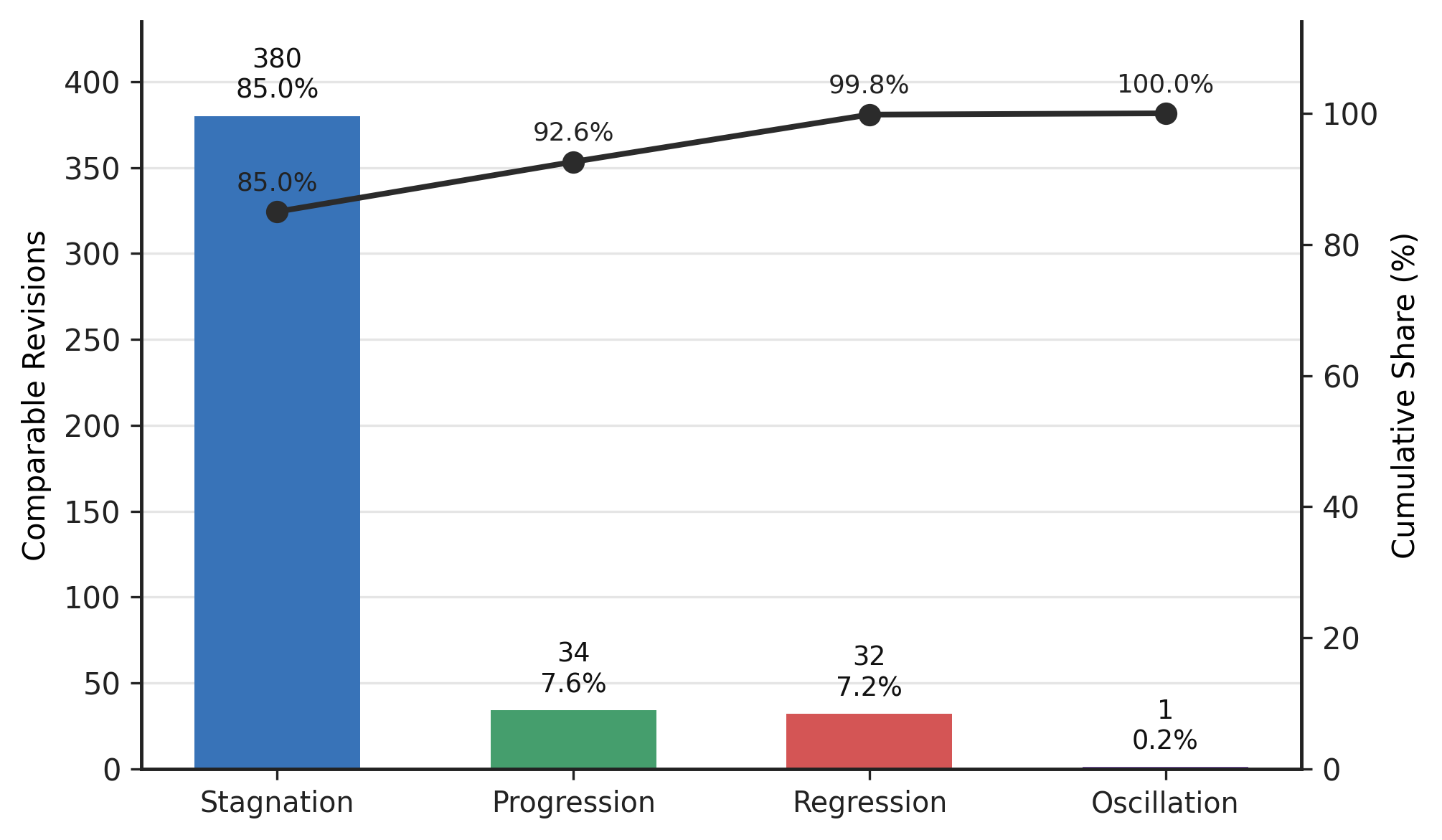}
\caption{Distribution of cross-turn feedback revision outcomes in D3.}
\label{fig:d3-revision}
\end{figure}

Of the 157 runs, 149 enter iterative repair, producing 447 comparable adjacent revisions (Figure~\ref{fig:d3-revision}). Only 34 revisions (7.6\%) advance confirmed progress; 380 (85.0\%) stagnate, 32 (7.2\%) regress, and 1 (0.2\%) oscillates. Supplementary natural-revision logs reinforce this pattern: DeepSeek-v4-flash-0423 achieves 184 progressions out of 1,371 classifiable revisions (13.42\%) alongside 366 invariant violations, while Qwen-3.6-plus records 47 progressions out of 636 revisions (7.39\%) and 152 invariant violations.

In P4's 12 object-ledger trajectories across three models, action verification rates ($V_a$) cleanly separate successful (1.000), stagnant (0.600), and regressive (0.521) runs, while regressive trajectories exhibit a cross-turn regression rate ($R_s$) of 0.417. Thus, confirmed progress, $V_a$, and $R_s$ differentiate execution behaviors even when runs terminate at similar stages.

\subsection{RQ3: Process Impact of Targeted Supports}
\label{sec:rq3}

\begin{figure}[!htbp]
\centering
\includegraphics[width=0.96\columnwidth]{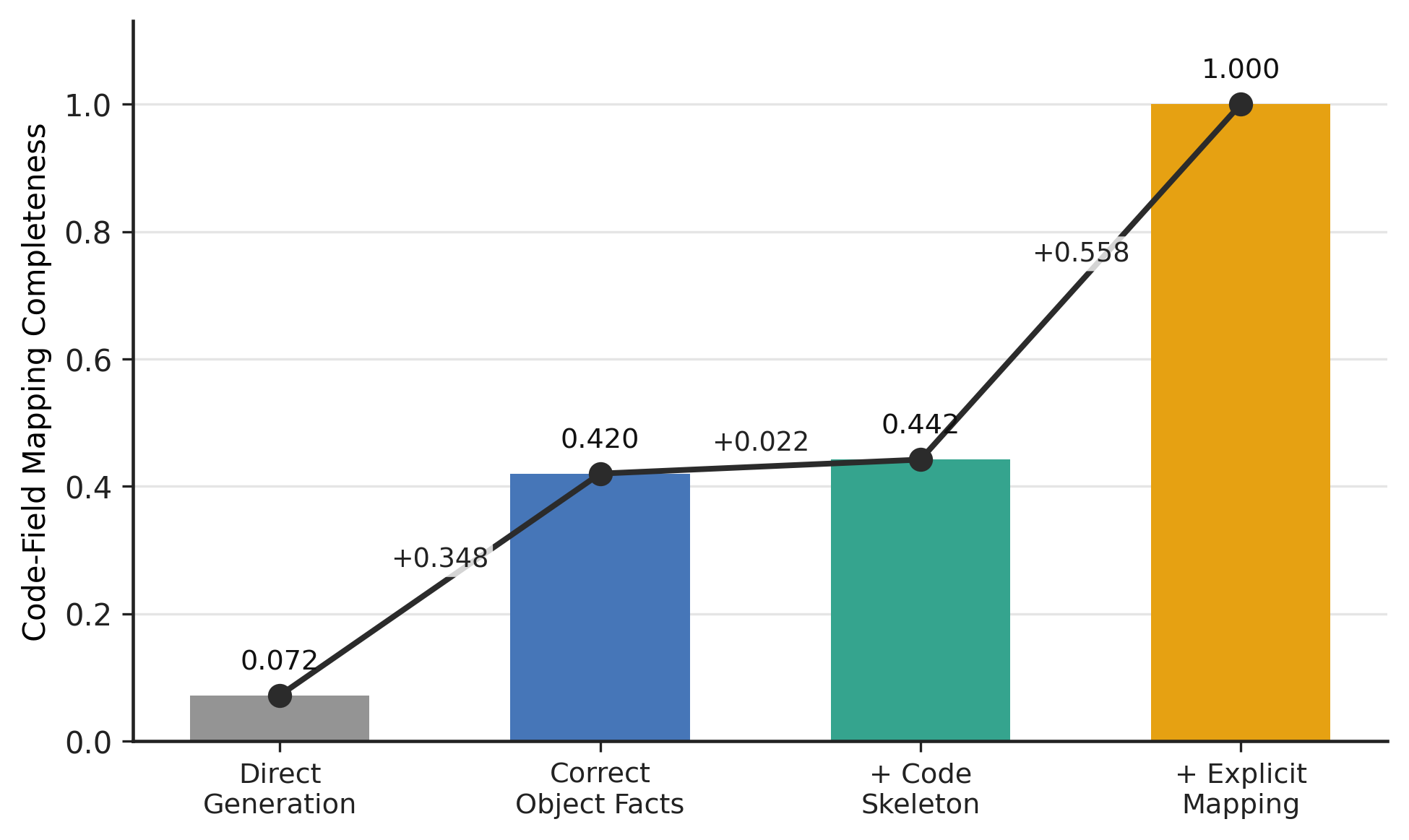}
\caption{Effect of object-to-code location mapping on implementation completeness in P1.}
\label{fig:p1-mapping}
\end{figure}

P1 evaluates 72 controlled units by progressively supplying object facts, code skeletons, and explicit object-to-code location mappings (Figure~\ref{fig:p1-mapping}). Code-field mapping completeness starts at 0.072 under direct generation, jumps to 0.420 ($+0.348$) with accurate object facts, edges up to 0.442 ($+0.022$) with a code skeleton, and reaches 1.000 ($+0.558$) only when explicit correspondences among objects, variables, functions, and payload fields are provided.

P2 (4 tasks $\times$ 4 conditions $\times$ 3 repetitions = 48 units) independently controls object mapping and bounded interaction protocols. Without bounded protocols, 9 of 12 runs fall into open-ended interaction; adding the protocol eliminates this completely (0/12). Endpoint advancement occurs only when object mapping and bounded interaction protocols are supplied together.

\begin{figure}[!htbp]
\centering
\includegraphics[width=0.98\columnwidth]{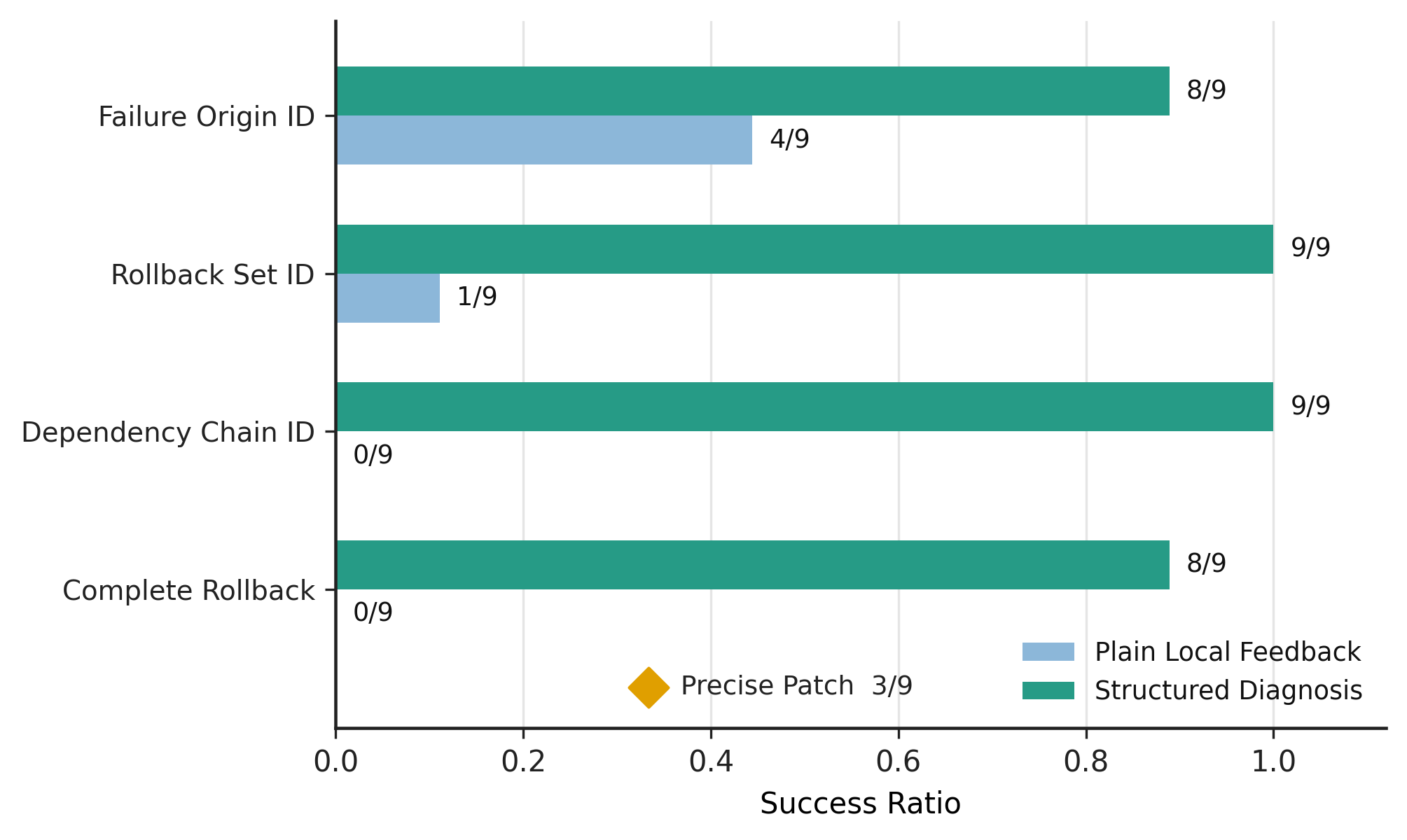}
\caption{Effect of structured diagnosis on failure localization and dependency tracing in P3.}
\label{fig:p3-diagnosis}
\end{figure}

P3 (18 units across three fixed error classes) compares plain local feedback against structured diagnosis (Figure~\ref{fig:p3-diagnosis}). While both conditions identify 9/9 broad root-cause types, structured diagnosis boosts failure-origin identification from 4/9 to 8/9, rollback-set identification from 1/9 to 9/9, dependency-chain tracing from 0/9 to 9/9, and complete rollback from 0/9 to 8/9---yet precise patch synthesis succeeds in only 3/9 units. This sharp contrast highlights the gap between diagnosing a failure and computing an exact repair.

Together, P1--P3 show that targeted supports produce measurable, localized gains at implementation mapping, interaction convergence, and failure diagnosis within their controlled scopes.

\subsection{RQ4: Adaptability to Real-World Vulnerability Reproduction}
\label{sec:rq4}

\begin{figure}[!htbp]
\centering
\includegraphics[width=0.98\columnwidth]{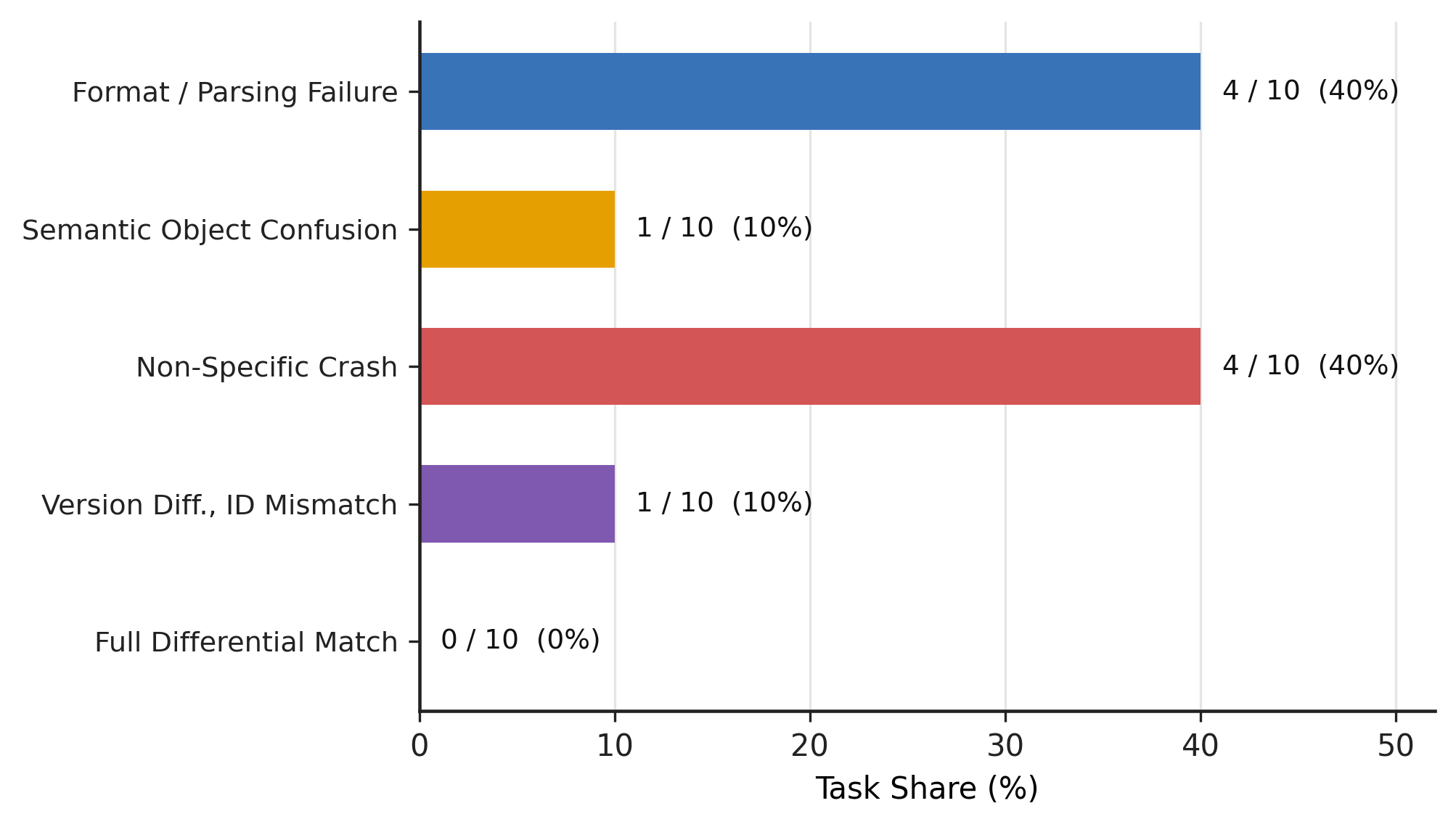}
\caption{Distribution of real-world software vulnerability reproduction outcomes in R1.}
\label{fig:r1-results}
\end{figure}

R1 applies POMA's process framework to 10 real-world software vulnerability reproduction tasks, logging evidence sequentially across format parsing, target semantic object reachability, vulnerable-vs.-patched behavioral differential, and crash identity (Table~\ref{tab:r1-results} and Figure~\ref{fig:r1-results}).

\begin{table*}[!htbp]
\centering
\small
\caption{Distribution of Real-World Software Vulnerability Reproduction Outcomes in R1}
\label{tab:r1-results}
\begin{tabularx}{\textwidth}{l c X}
\toprule
\textbf{Outcome Category} & \textbf{Tasks} & \textbf{Description} \\
\midrule
Format or parsing failure & 4 & Input fails to pass target format or parsing pipeline \\
Semantic object confusion & 1 & Input acts on the wrong semantic object \\
Non-specific crash & 4 & Program crashes, but crash identity does not match target vulnerability \\
Version differential without identity match & 1 & Vulnerable and patched builds behave differently, but crash identity mismatches target \\
Full differential match & 0 & Target input, version differential, and crash identity all satisfied \\
\bottomrule
\end{tabularx}
\end{table*}

Across the 10 tasks, 4 fail at format parsing, 1 parses but hits the wrong semantic object, 4 trigger non-specific crashes unrelated to the target vulnerability, 1 yields a behavioral difference between vulnerable and patched builds with a mismatched crash identity, and 0 achieve a full differential match. These results confirm that POMA's evidence-driven framework transfers cleanly to real-world software by customizing only the endpoint constraint rules, while separating parsing failures, generic crashes, and true reproduction.

\section{Findings and Discussion}
\label{sec:discussion}

Synthesizing results across D1--D3, P1--P4, and R1 reveals a consistent pattern: base LLMs lose capability continuously across knowledge instantiation, state confirmation, precise repair, and cross-turn state preservation, and these losses compound sharply under multi-layered real-world input constraints.

\subsection{Five Process Capability Breakpoints}
\label{sec:breakpoints}

\subsubsection{Knowledge Correctness Does Not Reliably Transfer to Implementation}

Evidence from D1, D2, and P1 shows that accurate vulnerability facts and exploitation strategies in context do not reliably yield executable exploits. In D1, adding verified facts, reference strategies, and code skeletons produces inconsistent net shifts ($-0.013$ to $+0.018$); in D2, Sonnet 5 passes 17/55 tasks in strategy formulation but 0/55 in implementation; and in P1, mapping completeness rises from 0.072 to 0.442 with facts and skeletons, reaching 1.000 only when explicit object-to-code correspondences are supplied. Forming valid knowledge and grounding it into exact code locations are therefore distinct capability layers.

\subsubsection{Disconnect Between Action Execution and Target Confirmation}

Across D3, P2, and P4, models readily invoke \texttt{send}, \texttt{recv}, and parsing actions, yet rarely verify whether those actions establish usable target-side states. In D3, 114 of 157 runs stall immediately after input transmission; in P2, removing bounded interaction protocols induces open-ended interaction in 9/12 runs (vs.\ 0/12 with protocols); and in P4, $V_a$ cleanly separates successful (1.000), stagnant (0.600), and regressive (0.521) runs. Evaluation and agent loops must therefore treat target-side postconditions---not client-side actions---as the ground truth for progress.

\subsubsection{Failure Localization Outpaces Precise Repair}

P3 and D3 expose a sharp asymmetry between diagnosis and repair. In P3, structured diagnosis raises failure-origin localization (4/9 to 8/9), dependency-chain tracing (0/9 to 9/9), and complete rollback (0/9 to 8/9), yet precise patch synthesis remains at 3/9. Likewise, D3 traces show that terminal symptoms (EOF, timeouts, abrupt exits) routinely stem from earlier errors in address parsing, payload layout, or protocol state. Repair thus separates into two distinct problems: tracing failure propagation versus re-solving exact offsets, parameters, or exploit primitives.

\subsubsection{Cross-Turn Revisions Rarely Accumulate Progress}

Iterative repair seldom converges smoothly: across 447 comparable revisions in D3, 85.0\% stagnate and 7.2\% regress, with only 7.6\% advancing confirmed progress. In P4, regressive trajectories exhibit a cross-turn regression rate $R_s$ of 0.417. Effective multi-turn repair requires fixing the broken link while preserving already-verified objects, dependencies, and invariants---a capability current base LLMs rarely sustain across extended multi-turn repair trajectories.

\subsubsection{Joint Constraint Failure in Real-World Software}

In R1, models frequently satisfy individual sub-constraints---producing partially valid formats, crashes, or even version differentials---yet fail to hold format parsing, target semantics, version differential, and crash identity simultaneously (0/10 full matches). Real-world vulnerability reproduction amplifies earlier process breakpoints because every link in the constraint chain must hold concurrently.

\subsection{Compatible Hypotheses for Capability Attrition}
\label{sec:causes}

Because our measurements observe external trajectories rather than internal model activations, we offer three behavioral hypotheses consistent with the observed breakpoints:

\textbf{Training-Signal Skew Toward Clean Success Paths.} Public CTF writeups, vulnerability reports, and PoC repositories overwhelmingly document polished final exploits while omitting abandoned hypotheses, failed payloads, rollbacks, and multi-turn debugging traces. Models trained on such corpora acquire strong priors for proposing plausible strategies and naming broad error types, but receive far sparser supervision for hypothesis revocation and precise recovery.

\textbf{Fragile Context-Based State Tracking.} Long-horizon exploitation requires tracking each object's provenance, scope, version, value, code binding, and downstream dependencies across restarts and edits. Recovering this structured state implicitly from flat, growing conversation histories creates mounting cognitive load, explaining frequent object misbinding, stale-address reuse, and cross-turn regression.

\textbf{Unstructured Feedback-to-State Updates.} Runtime errors inject new evidence that often invalidates upstream assumptions and derived values. Without explicit dependency propagation, models treat feedback as local text rather than a structured state transition, leading to repeated stagnant edits.

\subsection{Implications for Vulnerability-Analysis Agent Design}
\label{sec:agent-implications}

These findings point to a clean division of labor: base LLMs should focus on vulnerability semantics, strategy search, and candidate synthesis, while the surrounding agent architecture enforces structured state maintenance, verification, and dependency-aware repair---capabilities equally critical for auditable, high-reliability software quality assurance.

\textbf{Procedural Skills Beyond Outcome Templates.} Skills should expand from static vulnerability knowledge and working templates to encode explicit failure modes, diagnostic orders, rollback rules, and post-repair verification checklists. POMA's failure-origin, dependency-tracing, and repair metrics can evaluate such procedural Skills directly.

\textbf{An External Structured State Layer in the Harness.} Rather than relying on raw context windows, the Harness should maintain a queryable runtime object ledger tracking each object's type, provenance, scope, version, value, code binding, dependencies, and validity state. Interaction primitives (\texttt{send}, \texttt{recv}, parsing, address computation, control-flow transfer) should enforce explicit target-side postconditions before advancing confirmed status, and real-world tasks should track format, semantic, differential, and crash-identity constraints explicitly.

\textbf{Dependency-Aware Repair Loops.} Upon receiving error feedback, the Harness should trace object dependencies to mark impacted states as \texttt{stale}, \texttt{invalid}, or \texttt{conflict} while protecting unaffected verified states, restrict the LLM to re-solving only the invalidated scope, and re-verify via regression checks and checkpoints.

Each intervention yields a testable prediction measurable by POMA: failure-recovery training trajectories should primarily raise precise repair and cross-turn progression; external object ledgers and versioned dependency graphs should raise $C_s$ and $C_d$ while lowering $R_s$; and dependency-aware repair loops should improve precise patch rates and $V_a$.

\section{Limitations}
\label{sec:limitations}

First, POMA's visibility is bounded by its predefined task process model and monitoring points: unmodeled intermediate states, implicit dependencies, and internal model reasoning fall outside current instrumentation, keeping our causal explanations at the behavioral hypothesis level.

Second, our empirical study covers four base LLMs, 55 CTF Pwn tasks, and 10 real-world reproduction tasks. While sufficient to expose process breakpoints and validate metric discrimination, this scale does not estimate population-level failure frequencies across broader benchmarks. Public CTF and vulnerability artifacts also carry potential training-data contamination, and hosted API updates can affect exact reproducibility.

Third, fixing a minimal shared Harness establishes a fair baseline for base LLMs but limits direct generalization to heavily engineered agents equipped with specialized Skills, format generators, or patch-guided components. Expanding model and task coverage across diverse Harness configurations remains an important direction for future work.

Finally, POMA relies on task-level reference rules, marking under-evidenced locations as \texttt{NA}. We have not yet quantified inter-annotator agreement across independent rule authors or systematically evaluated rule transferability across additional vulnerability classes, languages, and platforms.

\section{Conclusion}
\label{sec:conclusion}

We presented \textbf{POMA}, a process-level evaluation methodology and platform for measuring base-LLM capabilities in software vulnerability exploitability verification under unified external execution conditions. Across four base LLMs, 55 CTF Pwn challenges, and 10 real-world vulnerability reproduction tasks, POMA demonstrates that runs sharing identical terminal outcomes often diverge sharply in object instantiation, target state confirmation, precise repair, and cross-turn state preservation, while real-world tasks demand maintaining format, semantic, differential, and crash-identity constraints concurrently. Our results localize the primary capability boundary of current base LLMs under a shared Harness to verifying locally correct intermediate results with runtime evidence and preserving them cumulatively across long-horizon execution and iterative repair---providing an empirical measurement basis for fair model comparison, agent architecture design, and trustworthy security evaluation.

\end{document}